\documentclass[10pt,twocolumn,letterpaper]{article}

\usepackage{wacv}
\usepackage{times}
\usepackage{epsfig}
\usepackage{graphicx}
\usepackage{amsmath}
\usepackage{amssymb}
\usepackage{multirow}
\usepackage{booktabs}
\usepackage[accsupp]{axessibility}
\usepackage{adjustbox}
\usepackage{xcolor}

\def\wacvPaperID{505} 

\wacvapplicationstrack 

\wacvfinalcopy 

\ifwacvfinal
\usepackage[breaklinks=true,bookmarks=false]{hyperref}
\else
\usepackage[pagebackref=true,breaklinks=true,colorlinks,bookmarks=false]{hyperref}
\fi

\begin{document}

\title{G-CASCADE: Efficient Cascaded Graph Convolutional Decoding for 2D Medical Image Segmentation}
\author{Md Mostafijur Rahman \qquad Radu Marculescu\\
The University of Texas at Austin\\
{\tt\small \{mostafijur.rahman, radum\}@utexas.edu}
}

\maketitle
\thispagestyle{empty}

\begin{abstract}
   In recent years, medical image segmentation has become an important application in the field of computer-aided diagnosis. In this paper, we are the first to propose a new graph convolution-based decoder namely, Cascaded Graph Convolutional Attention Decoder (G-CASCADE), for 2D medical image segmentation. G-CASCADE progressively refines multi-stage feature maps generated by hierarchical transformer encoders with an efficient graph convolution block. The encoder utilizes the self-attention mechanism to capture long-range dependencies, while the decoder refines the feature maps preserving long-range information due to the global receptive fields of the graph convolution block. Rigorous evaluations of our decoder with multiple transformer encoders on \color{black}five \color{black}medical image segmentation tasks (i.e., Abdomen organs, Cardiac organs, Polyp lesions, Skin lesions, \color{black}and Retinal vessels\color{black}) show that our model outperforms other state-of-the-art (SOTA) methods. We also demonstrate that our decoder achieves better DICE scores than the SOTA CASCADE decoder \color{black}with 80.8\% fewer parameters and 82.3\% fewer FLOPs\color{black}. Our decoder can easily be used with other hierarchical encoders for general-purpose semantic and medical image segmentation tasks.
   

\end{abstract}

\section{Introduction}

Automatic medical image segmentation plays a crucial role in the diagnosis, treatment planning, and post-treatment evaluation of various diseases; this involves classifying pixels and generating segmentation maps to identify lesions, tumours, or organs. Convolutional neural networks (CNNs) have been extensively utilized for medical image segmentation tasks \cite{ronneberger2015u, oktay2018attention, zhou2018unet++, huang2020unet, fan2020pranet, lou2021dc}. Among them, the U-shaped networks such as UNet \cite{ronneberger2015u}, UNet++ \cite{zhou2018unet++}, UNet 3+ \cite{huang2020unet}, and DC-UNet \cite{lou2021dc} exhibit reasonable performance 
and produce high-resolution segmentation maps. Additionally, researchers have incorporated attention modules into their architectures \cite{oktay2018attention, chen2018reverse, fan2020pranet} to enhance feature maps and improve pixel-level classification of medical images by capturing salient features. Although these attention-based methods have shown improved performance, they still struggle to capture long-range dependencies 
\cite{Rahman_2023_WACV}.

Recently, vision transformers \cite{dosovitskiy2020image} has shown great promise in capturing long-range dependencies among pixels and demonstrated improved performance, particularly for
medical image segmentation \cite{chen2021transunet, cao2021swin, dong2021polyp, wang2022stepwise, Rahman_2023_WACV, rahman2023multi, zhang2021transfuse, wang2022uctransnet}. The self-attention (SA) mechanism used in transformers learns correlations between input patches; this enables capturing the long-range dependencies among pixels. Recently, hierarchical vision transformers such as the Swin transformer \cite{liu2021swin}, 
the pyramid vision transformer (PVT) \cite{wang2021pyramid}, 
MaxViT \cite{tu2022maxvit}, 
MERIT \cite{rahman2023multi}, 
have been introduced to enhance performance. These hierarchical vision transformers are effective in medical image segmentation tasks \cite{chen2021transunet, cao2021swin, dong2021polyp, wang2022stepwise, Rahman_2023_WACV, rahman2023multi}. As self-attention modules employed in transformers have limited capacity to learn (local) spatial relationships among pixels \cite{chu2021conditional, islam2020much}, some methods \cite{xie2021segformer, wang2022uformer, wang2022pvt, dong2021polyp, wang2022stepwise, Rahman_2023_WACV, rahman2023multi} incorporate local convolutional attention modules in the decoder. However, due to the locality of convolution operations, these methods have difficulties at capturing long-range 
correlations among pixels.

To overcome the aforementioned limitations, we introduce a new Graph based CAScaded  Convolutional Attention DEcoder (G-CASCADE) using graph convolutions. More precisely, G-CASCADE enhances the feature maps by preserving long-range attention due to the global receptive field of the graph convolution operation, while incorporating local attention through the spatial attention mechanism. Our contributions are as follows:
\begin{itemize}
  \item \textbf{New Graph Convolutional Decoder:} We introduce a new graph-based cascaded convolutional attention decoder (G-CASCADE) for 2D medical image segmentation; this takes the multi-stage features of vision transformers and learns multiscale and multiresolution spatial representations. To the best of our knowledge, we are the first to propose this graph convolutional network-based decoder for semantic segmentation.
\item \textbf{Efficient Graph Convolutional Attention Block:} We introduce a new graph convolutional attention module to build our decoder; this preserves the long-range attention of the vision transformer and highlights salient features by suppressing irrelevant regions. The use of graph convolution makes our decoder efficient. 
\item \textbf{Efficient Design of Up-Convolution Block:} We design an efficient up-convolution block that enables computational gains without degrading performance.
\item \textbf{Improved Performance:} We empirically show that G-CASCADE can be used with any hierarchical vision encoder (e.g., PVT \cite{wang2022pvt}, MERIT \cite{chen2021transunet}) while significantly improving the performance of 2D medical image segmentation. When compared against multiple baselines, G-CASCADE produces better results than SOTA methods on ACDC, Synapse Multi-organ, ISIC2018 skin lesion, Polyp, \color{black}and Retinal vessels segmentation \color{black}benchmarks with a significantly lower computational cost.
\end{itemize}

The remaining of this paper is organized as follows: Section \ref{sec:related_work} summarizes the related work in vision transformers, graph convolutional networks, and medical image segmentation. Section \ref{sec:method} describes the proposed method 
Section \ref{sec:experiments} explains experimental setup and results on multiple medical image segmentation benchmarks. Section \ref{sec:ablation_study} covers different ablation experiments. Lastly, Section \ref{sec:conclusion} concludes the paper. 


\begin{figure*}[t]
\begin{center}
\includegraphics[width=0.9\linewidth]{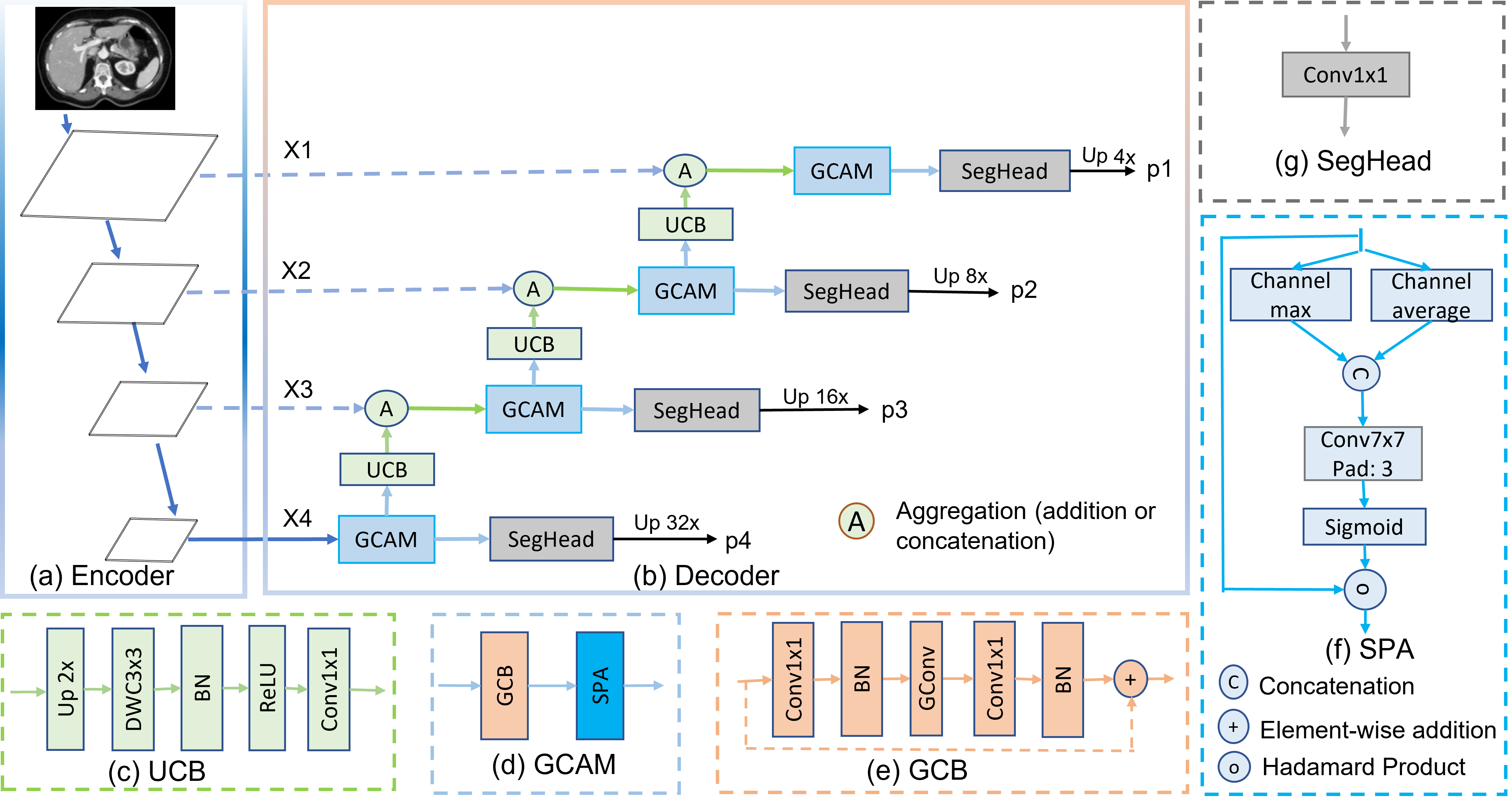}
\end{center}
   \caption{Hierarchical encoder with G-CASCADE network architecture. (a) PVTv2-b2 Encoder backbone with four stages, (b) G-CASCADE decoder, (c) Up-convolution block (UCB), (d) Graph convolutional attention module (GCAM), (e) Graph convolution block (GCB), (f) Spatial attention (SPA), and (g) Segmentation head (SegHead). X1, X2, X3, and X4 are the output features of the four stages of hierarchical encoder. p1, p2, p3, and p4 are output segmentation maps from four stages of our decoder.}
\label{fig:architecture}
\end{figure*}

\section{Related Work}
\label{sec:related_work}
We divide the related work into three parts, i.e., vision transformers, vision graph convolutional networks, and medical image segmentation; these are described next. 

\subsection{Vision transformers}
Dosovitskiy et al. \cite{dosovitskiy2020image} pioneered the development of the vision transformer (ViT), which enables the learning of long-range relationships between pixels through self-attention. Subsequent works have focused on enhancing ViT in various ways, such as integration of convolutional neural networks (CNNs) \cite{wang2022pvt,tu2022maxvit}, introducing new SA blocks \cite{liu2021swin,tu2022maxvit}, and novel architectural designs \cite{wang2021pyramid, xie2021segformer}. Liu et al. \cite{liu2021swin} introduce a sliding window attention mechanism within the hierarchical Swin transformer. Xie et al. \cite{xie2021segformer} present SegFormer, a hierarchical transformer utilizing Mix-FFN blocks. Wang et al. \cite{wang2021pyramid} develop the pyramid vision transformer (PVT) with a spatial reduction attention mechanism, and subsequently extend it to PVTv2 \cite{wang2022pvt} by incorporating overlapping patch embedding, a linear complexity attention layer, and a convolutional feed-forward network. Most recently, Tu et al. \cite{tu2022maxvit} introduce MaxViT, which employs a multi-axis self-attention mechanism to construct a hierarchical CNN-transformer encoder.

Although vision transformers exhibit remarkable performance, they have certain limitations in their (local) spatial information processing capabilities. In this paper, we aim to overcome these limitations by introducing a new graph-based cascaded attention decoder that preserves the long-range attention through graph convolution and incorporates local attention by a spatial attention mechanism.

\subsection{Vision graph convolutional networks}
Graph convolutional networks (GCNs) are developed primarily focusing on point clouds classification \cite{landrieu2018large, li2019deepgcns}, scene graph generation \cite{xu2017scene}, and action recognition \cite{yan2018spatial} in computer vision. Vision GNN (ViG) \cite{han2022vision} introduces the first graph convolutional backbone network to directly process the image data. ViG devides the image into patches and then uses K-nearest neighbors (KNN) algorithm to connect various patches; this enables the processing of long-range dependencies similar to vision transformers. Besides, due to using $1\times1$ convolutions before and after the graph convolution operation, the graph convolution block used in ViG is significantly faster than the vision transformer and $3\times3$ convolution-based CNN blocks. Therefore, we propose to use the graph convolution block to decode feature maps for dense prediction. This will make our decoder computationally efficient, while preserving long-range information.  

\subsection{Medical image segmentation}

Medical image segmentation is the task of classifying pixels into lesions, tumours, or organs in a medical image (e.g., endoscopy, MRI, and CT) \cite{chen2021transunet}. To address this task, U-shaped architectures \cite{ronneberger2015u, oktay2018attention, zhou2018unet++, huang2020unet, lou2021dc} have been commonly utilized due to their sophisticated encoder-decoder structure. Ronneberger et al. \cite{huang2020unet} introduce UNet, an encoder-decoder architecture that utilizes skip connections to aggregate features from multiple stages. In UNet++ \cite{zhou2018unet++}, nested encoder-decoder sub-networks are connected through dense skip connections. UNet 3+ \cite{huang2020unet} further extends this concept by exploring full-scale skip connections with intra-connections among the decoder blocks. DC-UNet \cite{lou2021dc} incorporates the multi-resolution convolution block and residual path within skip connections. These architectures have proven to be effective in medical image segmentation tasks.

Recently, transformers have gained popularity in the field of medical image segmentation \cite{cao2021swin, chen2021transunet, dong2021polyp, Rahman_2023_WACV, rahman2023multi, wang2022uctransnet, zhang2021transfuse}. In TransUNet \cite{chen2021transunet}, a hybrid architecture combining CNNs and transformers is proposed to capture both local and global pixel relationships. Swin-Unet \cite{cao2021swin} adopts a pure U-shaped transformer structure by utilizing Swin transformer blocks \cite{liu2021swin} in both the encoder and decoder. 
More recently, Rahman et al. \cite{rahman2023multi} propose a multi-scale hierarchical transformer network with cascaded attention decoding (MERIT) that calculates self attention in varying window sizes to capture effective multi-scale features.


Attention mechanisms have also been explored in combination with both CNNs \cite{oktay2018attention, fan2020pranet} and transformer-based architectures \cite{dong2021polyp} in medical image segmentation. PraNet \cite{fan2020pranet} utilizes the reverse attention mechanism \cite{chen2018reverse}. In PolypPVT \cite{dong2021polyp}, authors employ PVTv2 \cite{wang2022pvt} as the encoder and integrates CBAM \cite{woo2018cbam} attention blocks in the decoder, along with other modules. CASCADE \cite{Rahman_2023_WACV} proposes a cascaded decoder that utilizes both channel attention \cite{hu2018squeeze} and spatial attention \cite{chen2017sca} modules for feature refinement. CASCADE extracts features from four stages of the transformer encoder and uses cascaded refinement to generate high-resolution segmentation maps.
Due to incorporating local information with global information of transformers, CASCADE exhibits remarkable performance in medical image segmentation. However, CASCADE decoder has two major limitations: this can lead to i) long-range attention deficit due using only convolution operations during decoding and ii) high computational inefficiency due to using three $3\times3$ convolutions in each stage of the decoder. We propose to use graph convolution to overcome these limitations. 




\section{Method}
\label{sec:method}

In this section, we first introduce a new G-CASCADE decoder, then explain two different transformer-based architectures (i.e., PVT-GCASCADE and MERIT-GCASCADE) incorporating our proposed decoder.

\subsection{Cascaded Graph Convolutional Decoder (G-CASCADE)}
Existing transformer-based models have limited (local) contextual information processing ability among pixels. As a result, the transformer-based model faces difficulties in locating the more discriminating local features. To address this issue, some works \cite{dong2021polyp,Rahman_2023_WACV, rahman2023multi} utilize computationally expensive 2D convolution blocks in the decoder. Although the convolution block helps to incorporate the local information, it results in long-range attention deficits. To overcome this problem, we propose a new cascaded graph convolutional decoder, G-CASCADE, for pyramid encoders.

As shown in Figure \ref{fig:architecture}(b), G-CASCADE consists of efficient up-convolution blocks (UCBs) to upsample the features, graph convolutional attention modules (GCAMs) to robustly enhance the feature maps, and segmentation heads (SegHeads) to get the segmentation output. We have four GCAMs for the four stages of pyramid features from the encoder. To aggregate the multi-scale features, we first aggregate (e.g., addition or concatenation) the upsampled features from the previous decoder block with the features from the skip connections. Afterward, we process the concatenated features using our GCAM for enhancing semantic information. We then send the output from each GCAM to a prediction head. Finally, we aggregate four different prediction maps to produce the final segmentation output.

\subsubsection{Graph convolutional attention module (GCAM)}
We use the graph convolutional attention modules to refine the feature maps. GCAM consists of a graph convolution block ($GCB$(.)) to refine the features preserving long-range attention and a spatial attention \cite{chen2017sca} ($SPA$(·)) block to capture the local contextual information as in Equation \ref{eq:gcam}:
\begin{equation}
    GCAM(x) = SPA(GCB(x))
    \label{eq:gcam}
\end{equation}
where $x$ is the input tensor and $GCAM$(·) represents the convolutional attention module. Due to using graph convolution, our GCAM is significantly more efficient than the convolutional attention module (CAM) proposed in \cite{Rahman_2023_WACV}.


\textbf{Graph Convolution Block (GCB):} The GCB is used to enhance the features generated using our cascaded expanding path. In our GCB, we follow the Grapher design of Vision GNN \cite{han2022vision}. GCB consists of a graph convolution layer $GConv$(.) and two $1 \times 1$ convolution layers $C$(·) each followed by a batch normalization layer $BN$(·) and a ReLU activation layer $R$(.). $GCB$(·) is formulated as Equation \ref{eq:conv_block}:\begin{equation}
    GCB(x) = R(BN(C(GConv(R(BN(C(x)))))))
    \label{eq:conv_block}
\end{equation}
where $GConv$ can be formulated using Equation \ref{eq:gconv}:
\begin{equation}
    GConv(x) = GELU(BN(DynConv(x)))
    \label{eq:gconv}
\end{equation}
where $DynConv$(.) is a graph convolution (e.g., max-relative, edge, GraphSAGE, and GIN) in dense dilated K-nearest neighbour (KNN) graph. $BN$(.) and $GELU$(.) are batch normalization and GELU activation, respectively.


\textbf{SPatial Attention (SPA):} The SPA determines \textit{where} to focus in a feature map; then it enhances those features. The spatial attention is formulated as Equation \ref{eq:sa}:
\color{black}
\small
\begin{equation}
    SPA(x) = Sigmoid(Conv([C_{max}(x), C_{avg}(x)])) \circledast x
    \label{eq:sa}
\end{equation}
\normalsize \color{black}
where $Sigmoid$(·) is a Sigmoid activation function. $C_{max}$(·) and $C_{avg}$(·) represent the maximum and average values obtained along the channel dimension, respectively. $Conv$(·) is a $7 \times 7$ convolution layer with padding $3$ to enhance local contextual information (as in \cite{dong2021polyp}). $\circledast$ is the Hadamard product.

\subsubsection{Up-convolution block (UCB)}
UCB progressively upsamples the features of the current layer to match the dimension to the next skip connection. Each UCB layer consists of an UpSampling $Up$(·) with scale-factor 2, a $3\times3$ depth-wise convolution $DWC$(·) with groups equal input channels, a batch normalization $BN$(·), a $ReLU$(.) activation, and a $1\times1$ convolution $Conv$(.). The $UCB$(·) can be formulated as Equation \ref{eq:up_conv}:
\begin{equation}
    UCB(x) = Conv(ReLU(BN(DWC(Up(x)))))
    \label{eq:up_conv}
\end{equation}
Our UCB is light-weight as we replace the $3\times3$ convolution with a depth-wise convolution after upsampling.  

\subsubsection{Segmentation head (SegHead)}
SegHead takes refined feature maps from the four stages of the decoder as input and predicts four output segmentation maps. Each SegHead layer consists of a $1\times1$ convolution $Conv_{1\times1}$(·) which takes feature maps having $N_i$ channels ($N_i$ is the number of channels in the feature map of stage $i$) as input and gives output with channels equal to number of target classes for multi-class but $1$ channel for binary prediction. The $SegHead$(·) is formulated as Equation \ref{eq:seg_head}:
\begin{equation}
    SegHead(x) = Conv_{1\times1}(x)
    \label{eq:seg_head}
\end{equation}


\subsection{Overall architecture}

To ensure effective generalization and the ability to process multi-scale features in medical image segmentation, we integrate our proposed G-CASCADE decoder with two different hierarchical backbone encoder networks such as PVTv2 \cite{wang2022pvt} and MERIT \cite{rahman2023multi}. PVTv2 utilizes convolution operations instead of traditional transformer patch embedding modules to consistently capture spatial information. MERIT utilizes two MaxViT \cite{tu2022maxvit} encoders with varying window sizes for self-attention, thus enabling the capture of multi-scale features. 

By utilizing the PVTv2-b2 (Standard) encoder, we create the PVT-GCASCADE architecture. To adopt PVTv2-b2, we first extract the features (X1, X2, X3, and X4) from four layers and feed them (i.e., X4 in the upsample path and X3, X2, X1 in the skip connections) into our G-CASCADE decoder as shown in Figure \ref{fig:architecture}(a-b). Then, the G-CASCADE processes them and produces four prediction maps that correspond to the four stages of the encoder network.

Besides, we introduce the new MERIT-GCASCADE architecture by adopting the architectural design of the MERIT network. In the case of MERIT, we only replace their decoder with our proposed decoder and keep their hybrid CNN-transformer MaxViT \cite{tu2022maxvit} encoder networks. In our MERIT-GCASCADE architecture, we extract hierarchical feature maps from four stages of first encoder and then feed them to the corresponding decoder. Afterwards, we aggregate the feedback from final stage of the decoder to the input image and feed them to second encoder having different window sizes for self-attention. We extract feature maps from four stages of the second decoder and feed them to the second decoder. We send cascaded skip connections like MERIT \cite{rahman2023multi} to the second decoder. We get four output segmentation maps from the four stages of our second decoder. Finally, we aggregate the segmentation maps from the two decoders for four stages separately to produce four output segmentation maps. Our proposed decoder is designed to be adaptable and seamlessly integrates with other hierarchical backbone networks. 




\subsection{Multi-stage outputs and loss aggregation}
We get four output segmentation maps $p_1$, $p_2$, $p_3$, and $p_4$ from the four prediction heads for the four stages of our G-CASCADE decoder. 

\textbf{Output segmentation maps aggregation:} We compute the final segmentation output using additive aggregation as in Equation \ref{eq:output_aggregation}:
\begin{equation}
    seg\_output = \alpha p_1 + \beta p_2 + \gamma p_3 + \zeta p_4
\label{eq:output_aggregation}
\end{equation}
where $\alpha$, $\beta$, $\gamma$, and $\zeta$ are the weights of each prediction head. We set $\alpha$, $\beta$, $\gamma$, and $\zeta$ to 1.0 in all our experiments. We get the final prediction output by applying the Sigmoid activation for binary segmentation and Softmax activation for multi-class segmentation. 

\textbf{Loss aggregation:} Following MERIT \cite{rahman2023multi}, we use the combinatorial loss aggregation strategy, MUTATION in all our experiments. Therefore, we compute the loss for $2^n-1$ combinatrorial predictions synthesized from $n$ heads separately and then do a summation of them. We optimize this additive combinatorial loss during training.



\begin{table*}[]
\begin{center}
    {\small{
\begin{tabular}{lrrrrrrrrrrrr}
\toprule
\multirow{2}{*}{Architectures} & \multicolumn{3}{c}{Average}                                                                                     & \multicolumn{1}{l}{\multirow{2}{*}{Aorta}} & \multicolumn{1}{l}{\multirow{2}{*}{GB}} & \multicolumn{1}{l}{\multirow{2}{*}{KL}} & \multicolumn{1}{l}{\multirow{2}{*}{KR}} & \multicolumn{1}{l}{\multirow{2}{*}{Liver}} & \multicolumn{1}{l}{\multirow{2}{*}{PC}} & \multicolumn{1}{l}{\multirow{2}{*}{SP}} & \multicolumn{1}{l}{\multirow{2}{*}{SM}} \\
                               & \multicolumn{1}{l}{DICE$\uparrow$} & \multicolumn{1}{l}{HD95$\downarrow$} & \multicolumn{1}{l}{mIoU$\uparrow$} & \multicolumn{1}{l}{}                       & \multicolumn{1}{l}{}                    & \multicolumn{1}{l}{}                         & \multicolumn{1}{l}{}                         & \multicolumn{1}{l}{}                       & \multicolumn{1}{l}{}                    & \multicolumn{1}{l}{}                    & \multicolumn{1}{l}{}                    \\
\midrule
UNet \cite{ronneberger2015u}                   & 70.11                    & 44.69                    & 59.39                                        & 84.00                                      & 56.70                                   & 72.41                                        & 62.64                                        & 86.98                                      & 48.73                                   & 81.48                                   & 67.96                                   
\\
AttnUNet \cite{oktay2018attention}                   & 71.70                    & 34.47                    & 61.38                                   & 82.61                                      & 61.94                                   & 76.07                                        & 70.42                                        & 87.54                                      & 46.70                                   & 80.67                                   & 67.66                                   
\\
R50+UNet \cite{chen2021transunet}                   & 74.68                    & 36.87                    & $-$                                     & 84.18                                      & 62.84                                   & 79.19                                        & 71.29                                        & 93.35                                      & 48.23                                   & 84.41                                   & 73.92                                   
\\
R50+AttnUNet \cite{chen2021transunet}                   & 75.57                    & 36.97                    & $-$                                    & 55.92                                      & 63.91                                   & 79.20                                        & 72.71                                        & 93.56                                      & 49.37                                   & 87.19                                   & 74.95                                   
\\
SSFormerPVT \cite{wang2022stepwise}                   & 78.01                    & 25.72                    & 67.23                                     & 82.78                                      & 63.74                                   & 80.72                                        & 78.11                                        & 93.53                                      & 61.53                                   & 87.07                                   & 76.61                                   \\
PolypPVT \cite{dong2021polyp}                       & 78.08                    & 25.61                    & 67.43                           & 82.34                                      & 66.14                                   & 81.21                                        & 73.78                                        & 94.37                                      & 59.34                                   & 88.05                                   & 79.4                                    \\
TransUNet \cite{chen2021transunet}                     & 77.61                    & 26.9                     & 67.32                                  & 86.56                                      & 60.43                                   & 80.54                                        & 78.53                                        & 94.33                                      & 58.47                                   & 87.06                                   & 75.00                                      \\
SwinUNet \cite{cao2021swin}                      & 77.58                    & 27.32                    & 66.88                              & 81.76                                      & 65.95                                   & 82.32                                        & 79.22                                        & 93.73                                      & 53.81                                   & 88.04                                   & 75.79                                   \\
MT-UNet \cite{wang2022mixed}                      & 78.59                    & 26.59             & $-$       & 87.92                                      & 64.99                                   & 81.47                                        & 77.29                                        & 93.06                                     & 59.46                                  & 87.75                                   & 76.81                                  \\
MISSFormer \cite{huang2021missformer}                      & 81.96                    & 18.20             & $-$       & 86.99                                      &  68.65                                   &  85.21                                        & 82.00                                        & 94.41                                      & 65.67                                   & 91.92                                   & 80.81                                   \\ 
PVT-CASCADE \cite{Rahman_2023_WACV}                  & 81.06                    & 20.23                    & 70.88                                     & 83.01                                      & 70.59                                   & 82.23                                        & 80.37                                        & 94.08                                      & 64.43                                   & 90.1                                    & 83.69                                   \\
TransCASCADE \cite{Rahman_2023_WACV}                & 82.68                    & 17.34                    & 73.48                                         & 86.63                                      & 68.48                                   & 87.66                                        & 84.56                                        & 94.43                                     & 65.33                                   & 90.79                                   & 83.52                                   \\
Cascaded MERIT \cite{rahman2023multi}                   & 84.32                    & 14.27    & 75.44                & 86.67                                      & 72.63                                   & 87.71                                       & 84.62                                        & 95.02                                      & \textbf{70.74}                                   &  \textbf{91.98}                                    & \textbf{85.17}                                   \\
\midrule
PVT-GCASCADE (\textbf{Ours})                 & 83.28                   &   15.83                  &  73.91                         &   86.50                                    &  71.71                                  &   87.07                                      &   83.77                                    &  95.31                                   &   66.72                                &   90.84                                & 83.58
\\
MERIT-GCASCADE (\textbf{Ours})                 & \textbf{84.54}                   & \textbf{10.38}                    & \textbf{75.83}                           &  \textbf{88.05}                                     &    \textbf{74.81}                                & \textbf{88.01}                                        & \textbf{84.83}                                        & \textbf{95.38}                                      & 69.73                                   & 91.92                                   & 83.63
\\
\bottomrule 
\end{tabular}
}}
\end{center}
\caption{Results of Synapse Multi-organ segmentation. We report only DICE scores for individual organs. We get the results of UNet, AttnUNet, PolypPVT, SSFormerPVT, TransUNet, and SwinUNet from \cite{Rahman_2023_WACV}. We reproduce the results of Cascaded MERIT with a batch size of 6. $\uparrow$ ($\downarrow$) denotes the higher (lower) the better. G-CASCADE results are averaged over five runs. The best results are shown in bold.}
\label{tab:multi_organ_results}
\end{table*}

\section{Experimental Evaluation}
\label{sec:experiments}
In this section, we first describe the dataset and evaluation metrics followed by implementation details. Then, we conduct a comparative analysis between our proposed G-CASCADE decoder-based architectures and SOTA methods to highlight the superior performance of our approach.

\subsection{Datasets}
\label{ssec:dataset_metrics}
We present the description of Synapse Multi-organ and ACDC datasets below. \textbf{The description of ISIC2018, polyp, \color{black}and retinal vessels segmentation \color{black}datasets are available in supplementary materials (Section A).}

\textbf{Synapse Multi-organ dataset.} The Synapse Multi-organ dataset\footnote{\href{https://www.synapse.org/\#!Synapse:syn3193805/wiki/217789}{https://www.synapse.org/\#!Synapse:syn3193805/wiki/217789 }} contains 30 abdominal CT scans which have 3779 axial contrast-enhanced slices. Each CT scan has 85-198 slices of $512 \times 512$ pixels. 
Similar to TransUNet \cite{chen2021transunet}, we divide the dataset randomly into 18 scans for training (2212 axial slices) and 12 scans for validation. We segment only 8 abdominal organs, i.e., aorta, gallbladder (GB), left kidney (KL), right kidney (KR), liver, pancreas (PC), spleen (SP), and stomach (SM). 

\textbf{ACDC dataset.} The ACDC dataset\footnote{\href{https://www.creatis.insa-lyon.fr/Challenge/acdc/}{https://www.creatis.insa-lyon.fr/Challenge/acdc/}} contains 100 cardiac MRI scans each of which consists of three organs, right ventricle (RV), myocardium (Myo), and left ventricle (LV). Following TransUNet \cite{chen2021transunet}, we use 70 cases (1930 axial slices) for training, 10 for validation, and 20 for testing.

\subsection{Evaluation metrics} 
\label{assec:eval_metrics}

We use DICE, mIoU, and 95\% Hausdorff Distance (HD95) to evaluate performance on the Synapse Multi-organ dataset. However, for the ACDC dataset, we use only DICE score as an evaluation metrics. We use DICE and mIoU as the evaluation metrics in polyp segmentation and ISIC2018 datasets. The DICE score $DSC(Y, \hat{Y})$, $IoU(Y, \hat{Y})$, and HD95 distance $D_H(Y, \hat{Y})$ are calculated using Equations \ref{eq:dice}, \ref{eq:iou}, and \ref{eq:95hd}, respectively.
\begin{equation}\label{eq:dice}
DSC(Y, \hat{Y}) = \frac{2 \times \lvert Y \cap \hat{Y} \rvert}{\lvert Y \rvert + \lvert \hat{Y} \rvert}\times100
\end{equation}
\begin{equation}\label{eq:iou}
IoU(Y, \hat{Y}) = \frac{\lvert Y \cap \hat{Y} \rvert}{\lvert Y \cup \hat{Y} \rvert}\times100
\end{equation}

\begin{equation}\label{eq:95hd}
\small
D_H(Y, \hat{Y}) = \max \{\max_{y \in Y} \min_{\hat{y} \in \hat{Y}}d(y, \hat{y}), \{\max_{\hat{y} \in \hat{Y}} \min_{y \in Y}d(y, \hat{y})\}
\end{equation}
where $Y$ and $\hat{Y}$ are the ground truth and predicted segmentation map, respectively. 


\subsection{Implementation details}
\label{ssec:impl_details}
We use Pytorch 1.11.0 to implement our network and conduct experiments. We train all models on a single NVIDIA RTX A6000 GPU with 48GB of memory. We use the PVTv2-b2 and Small CascadedMERIT as representative network. We use the pre-trained weights on ImageNet for both PVT and MERIT backbone networks. We train our model using AdamW optimizer \cite{loshchilov2017decoupled} with both learning rate and weight decay of 0.0001. 

\textbf{GCB:} We construct dense dilated graph using $K=11$ neighbors for KNN and use the \textit{Max-Relative (MR)} graph convolution in all our experiments. The \textit{batch normalization} is used after MR graph convolution. Following ViG \cite{han2022vision}, we also use the relative position vector for graph construction and reduction ratios of [1, 1, 4, 2] for graph convolution block in different stages.   

\textbf{Synapse Multi-organ dataset.}
We use a batch size of 6 and train each model for maximum of 300 epochs. We use the input resolution of $224 \times 224$ for PVT-GCASCADE and ($256\times256$, $224\times224$) for MERIT-GCASCADE. We apply random rotation and flipping for data augmentation. The combined weighted Cross-entropy (0.3) and DICE (0.7) loss are utilized as the loss function.

\textbf{ACDC dataset.}
For the ACDC dataset, we train each model for a maximum of 150 epochs with a batch size of 12. We set the input resolution as $224 \times 224$ for PVT-GCASCADE and ($256\times256$, $224\times224$) for MERIT-GCASCADE. We apply random flipping and rotation for data augmentation. We optimize the combined weighted Cross-entropy (0.3) and DICE (0.7) loss function.

\textbf{ISIC2018 dataset:} We resize the images into $384\times384$ resolution. Then, we train our model for 200 epochs with a batch size of 4 and a gradient clip of 0.5. We optimize the combined weighted BCE and weighted IoU loss function.

\textbf{Polyp datasets.}
We resize the image to $352 \times 352$ and use a multi-scale \{0.75, 1.0, 1.25\} training strategy with a gradient clip limit of 0.5 like CASCADE \cite{Rahman_2023_WACV}. We use a batch size of 4 and train each model a maximum of 200 epochs. We optimize the combined weighted BCE and weighted IoU loss function.

\begin{table}[]
\begin{center}
\begin{adjustbox}{width=0.48\textwidth}
\begin{tabular}{lrrrr}
\toprule
\multirow{2}{*}{Methods}        & Avg     & \multicolumn{3}{l}{} \\
& Dice & \multicolumn{1}{l}{RV} & \multicolumn{1}{l}{Myo} & \multicolumn{1}{l}{LV} \\
\midrule
R50+UNet   \cite{chen2021transunet}       & 87.55                        & 87.10                  & 80.63                   & 94.92                  \\
R50+AttnUNet  \cite{chen2021transunet}  & 86.75                        & 87.58                  & 79.20                   & 93.47                  \\
ViT+CUP \cite{chen2021transunet}   & 81.45                        & 81.46                  & 70.71                   & 92.18                 \\
R50+ViT+CUP \cite{chen2021transunet} & 87.57                        & 86.07                  & 81.88                   & 94.75                  \\
TransUNet  \cite{chen2021transunet}       & 89.71                        & 86.67                  & 87.27                   & 95.18                  \\
SwinUNet \cite{cao2021swin}         & 88.07                        & 85.77                  & 84.42                   & 94.03                  \\
MT-UNet \cite{wang2022mixed}         & 90.43                        & 86.64                  & 89.04                   & 95.62                  \\
MISSFormer \cite{huang2021missformer}         & 90.86                        & 89.55                  & 88.04                   & 94.99                  \\
PVT-CASCADE \cite{Rahman_2023_WACV}      & 91.46                        & 89.97                   & 88.9                   & 95.50                   \\
TransCASCADE \cite{Rahman_2023_WACV}    & 91.63                       & 90.25                   &  89.14                  & 95.50 \\
Cascaded MERIT \cite{rahman2023multi}    & 91.85                       & 90.23                   &  89.53                  & 95.80 \\
\midrule
PVT-GCASCADE (\textbf{Ours})    & 91.95                        & 90.31                  & 89.63                   & 95.91 \\
MERIT-GCASCADE (\textbf{Ours})    & \textbf{92.23}                        & \textbf{90.64}                  & \textbf{89.96}                   & \textbf{96.08} \\
\bottomrule 
\end{tabular}
\end{adjustbox}
\end{center}
\caption{Results on ACDC dataset. DICE scores are reported for individual organs. We get the results of SwinUNet from \cite{Rahman_2023_WACV}. G-CASCADE results are averaged over five runs. The best results are shown in bold.} 
\label{tab:acdc_results}
\end{table}

\begin{figure*}[t]
\begin{center}
\includegraphics[width=0.9\linewidth]{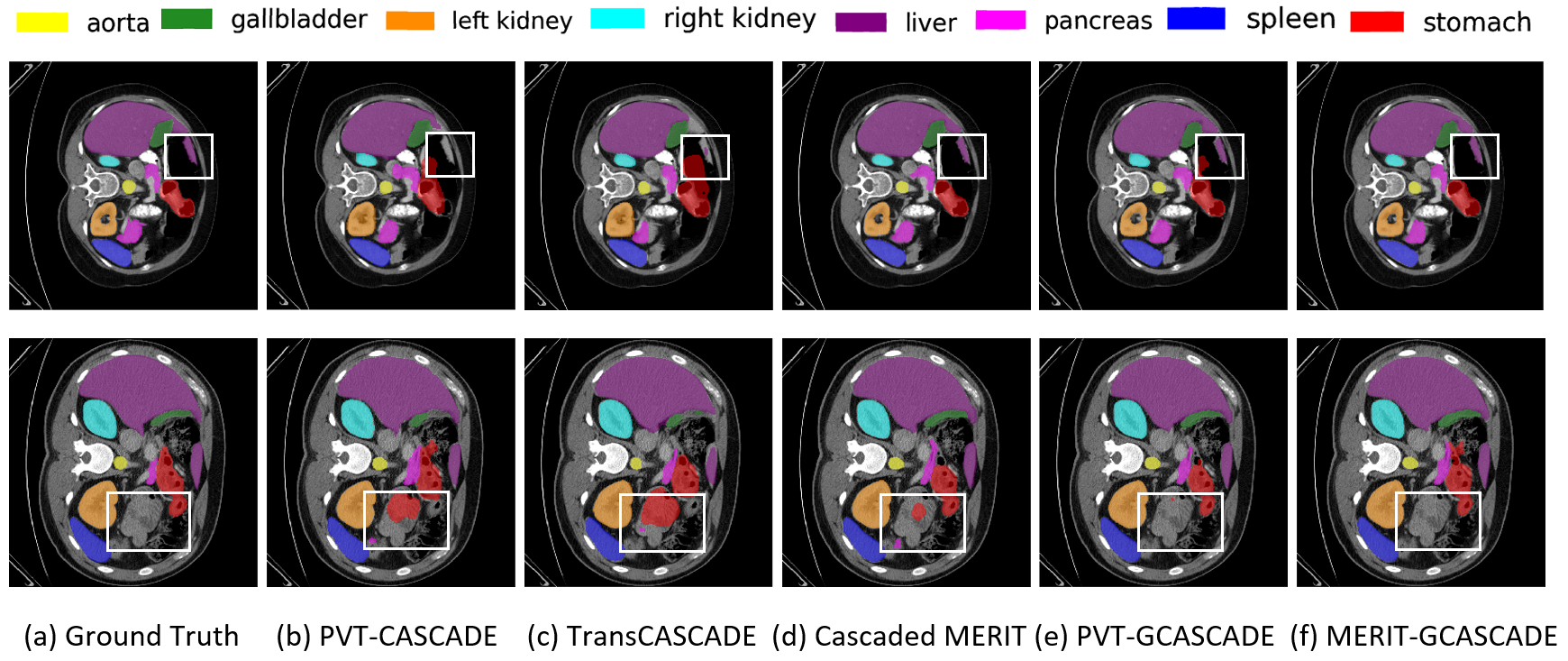}
\end{center}
   \caption{Qualitative results on Synapse multi-organ dataset. (a) Ground Truth (GT), (b) PVT-CASCADE, (c) TransCASCADE, (d) Cascaded MERIT, (e) PVT-GCASCADE, and (f) MERIT-GCASCADE. We overlay the segmentation maps on top of original image/slice. We use the white bounding box to highlight regions where most of the methods have incorrect predictions.}
\label{fig:qualitative_synapse}
\end{figure*}

\subsection{Results}
\label{ssec:results}
We compare our architectures (i.e., PVT-GCASCADE and MERIT-GCASCADE) with SOTA CNN and transformer-based segmentation methods
on Synapse Multi-organ, ACDC, ISIC2018 \cite{codella2019skin}, and Polyp (i.e., Endoscene  \cite{vazquez2017benchmark}, CVC-ClinicDB \cite{bernal2015wm}, Kvasir \cite{jha2020kvasir}, ColonDB \cite{tajbakhsh2015automated}) datasets. \textbf{The results of ISIC2018, polyp, \color{black}and retinal vessels segmentation \color{black}datasets are reported in the supplementary materials (Section B)}.

\subsubsection{Quantitative results on Synapse Multi-organ dataset}
Table \ref{tab:multi_organ_results} presents the performance of different CNN- and transformer-based methods on Synapse Multi-organ segmentation dataset. We can see from Table \ref{tab:multi_organ_results} that our MERIT-GCASCADE significantly outperforms all the SOTA CNN- and transformer-based 2D medical image segmentation methods thus achieving the best average DICE score of 84.54\%. Our PVT-GCASCADE and MERIT-GCASCADE outperforms their counterparts PVT-CASCADE and Cascaded MERIT by 2.22\% and 0.22\% DICE scores, respectively with significantly lower computational costs. Similarly, our PVT-GCASCADE and MERIT-GCASCADE outperforms their counterparts by 4.4 and 3.89 in HD95 distance. Our MERIT-GCASCADE has the lowest HD95 distance (10.38) which is 3.89 lower than the best SOTA method Cascaded MERIT (HD95 of 14.27). The lower HD95 scores indicate that our G-CASCADE decoder can better locate the boundary of organs.

Our proposed decoder also shows boost in the DICE scores of individual organ segmentation. We can see from the Table \ref{tab:multi_organ_results} that our proposed MERIT-GCASCADE significantly outperforms SOTA methods on five out of eight organs. We believe that G-CASCADE decoder demonstrates better performance due to using graph convolution together with the transformer encoder.

\subsubsection{Quantitative results on ACDC dataset}
We have conducted another set of experiments on the MRI images of the ACDC dataset using our architectures. Table \ref{tab:acdc_results}
presents the average DICE scores of our PVT-GCASCADE and MERIT-GCASCADE along with other SOTA methods. Our MERIT-GCASCADE achieves the highest average DICE score of 92.23\% thus improving about 0.38\% over Cascaded MERIT though our decoder has significantly lower computational cost (see Table \ref{tab:compare_baseline_decoder}). Our PVT-GCASCADE gains 91.95\% DICE score which is also better than all other methods. Besides, both our PVT-GCASCADE and MERIT-GCASCADE have better DICE scores in all three organs segmentation. 

\subsubsection{Qualitative results on Synapse Multi-organ dataset}
\label{ssec:qualitative_synapse}
We present the segmentation outputs of our proposed method and three other SOTA methods on two sample images in Figure \ref{fig:qualitative_synapse}.
If we look into the highlighted regions in both samples, we can see that MERIT-GCASCADE consistently segments the organs with minimal false negative and false positive results. PVT-GCASCADE and Cascaded MERIT show comparable results. PVT-GCASCADE has false positives in first sample (i.e., first row) and has better segmentation in second sample (i.e., second row), whereas Cascaded MERIT provides better segmentation in first sample but it has larger false positives in second sample. TransCASCADE and PVT-CASCADE provide larger incorrect segmentation outputs in both samples.

\begin{table}[]
\begin{center}
    \begin{adjustbox}{width=0.48\textwidth}
\begin{tabular}{cccrrr}
\toprule
\multicolumn{3}{c}{Components} & \multicolumn{1}{c}{FLOPs} & \multicolumn{1}{c}{\#Params} & \multicolumn{1}{r}{Avg}  \\
Cascaded     & GCB     & SPA   & \multicolumn{1}{c}{(G)} & \multicolumn{1}{c}{(M)} & DICE                    \\
\midrule
No           & No     & No     &   0 & 0 & 80.1$\pm$0.2     \\
Yes          & No     & No     & 0.102 & 0.225 & 81.1$\pm$0.2    \\
Yes          & No    & Yes     & 0.102 & 0.225 & 82.1$\pm$0.3          \\
Yes          & Yes     & No    & 0.341 & 1.78 & 83.0$\pm$0.2      \\
Yes          & Yes    & Yes    & 0.342 & 1.78 & \textbf{83.3$\pm$0.2}       \\
\bottomrule 
\end{tabular}
\end{adjustbox}
\end{center}
\caption{Quantitative results of different components of G-CASCADE with PVTv2-b2 encoder on Synapse multi-organ dataset. We use \textit{additive aggregation} for adding skip connections and an input resolution of $224\times224$ to get these results. All results are averaged over five runs. The best results are showed in bold.}
\label{tab:ablation_components}
\end{table}

\begin{table}[t]
\centering 
    {
{\begin{tabular}{lrrrr}
\toprule
Arrangements  & \multicolumn{1}{l}{DICE (\%)} \\
\midrule
SPA $\rightarrow$ GCB & 82.93$\pm$0.2  
\\ GCB $\rightarrow$ SPA (\textbf{Ours}) & \textbf{83.28$\pm$0.2}  \\
\bottomrule \\
\end{tabular}}

}
\caption{Comparison of different arrangements of GCB and SPA in GCAM on Synapse Multi-organ dataset. We use PVTv2-b2 as the encoder to produce these results. All the results are averaged over five runs. The best results are in bold.}
\label{tab:arrangements_sa_gcb}
\end{table}

\section{Ablation Study}
\label{sec:ablation_study}

In this section, we perform a set of ablation experiments that aim to address various questions concerning our proposed architectures and experimental setup. \textbf{More ablation studies are available in supplementary materials (Section C).}



\subsection{Effect of different components of G-CASCADE}
We carry out ablation studies on the Synapse Multi-organ dataset to evaluate the effectiveness of different components of our proposed G-CASCADE decoder. We use the same PVTv2-b2 backbone pre-trained on ImageNet and the same experimental settings for Synapse Multi-organ dataset in all experiments. We remove different modules such as Cascaded structure, GCB, and SPA from the G-CASCADE decoder and compare the results. It is evident from Table \ref{tab:ablation_components} that the cascaded structure of the decoder improves performance over the non-cascaded decoder. GCB and SPA modules also help improve performance. However, the use of both SPA and GCB modules together produces the best DICE score of 83.3\%. We can also see from the table that DICE score is improved about 3.2\% with 0.342G and 1.78M additional FLOPs and parameters, respectively.

\subsection{Effect of arrangements of GCB and SPA in GCAM}
We have conducted an ablation study to see the effect of the order of GCB and SPA in GCAM. Table \ref{tab:arrangements_sa_gcb} presents the experimental results of two different arrangements. We can conclude from Table \ref{tab:arrangements_sa_gcb} that GCB followed by SPA block performs better than SPA followed by GCB. Therefore, in our G-CASCADE decoder, we use a GCB followed by a SPA block in each GCAM.
\begin{table}[t]
\centering 
    {
\begin{adjustbox}{width=0.48\textwidth}
{\begin{tabular}{lrrrr}
\toprule
Decoders    & UCB &\multicolumn{1}{c}{FLOPs(G)} & \multicolumn{1}{c}{\#Params(M)} & \multicolumn{1}{l}{DICE (\%)} \\
\midrule
CASCADE         & Original & 1.93 & 9.27 & 82.78  \\
CASCADE       & Modified    & 1.22 & 7.58 & 82.79   \\
\midrule 
G-CASCADE (\textbf{Ours})       & Original & 1.06 & 3.47 & 83.15      \\
G-CASCADE (\textbf{Ours})     & Modified & \textbf{0.342} & \textbf{1.78}     & \textbf{83.28}   \\
\bottomrule \\
\end{tabular}}
\end{adjustbox}

}
\caption{Comparison with the baseline decoder on Synapse Multi-organ dataset. We only report the FLOPs and the number of parameters of the respective decoder. We produce these results using PVTv2-b2 encoder. All the results are averaged over five runs. The best results are in bold.}
\label{tab:compare_baseline_decoder}
\end{table}

\begin{table}[t]
\centering 
    {
\begin{adjustbox}{width=0.48\textwidth}
{\begin{tabular}{lrrrr}
\toprule
Architectures    &  Aggregation   &  \multicolumn{1}{c}{FLOPs(G)} & \multicolumn{1}{c}{\#Params(M)} & \multicolumn{1}{l}{DICE (\%)} \\
\midrule
PVT-GCASCADE            & Addition            & \textbf{0.342} & \textbf{1.78} & 83.28  \\
PVT-GCASCADE         & Concat                    & 0.975 & 3.32 & \textbf{83.40}      \\
\midrule 
MERIT-GCASCADE           & Addition                            & \textbf{1.523} & \textbf{3.55} & 84.54   \\
MERIT-GCASCADE        & Concat               & 4.27 & 5.99     & \textbf{84.63}   \\
\bottomrule \\
\end{tabular}}
\end{adjustbox}

}
\caption{Comparison of different skip-aggregations in G-CASCADE decoder on Synapse Multi-organ dataset. We only report the FLOPs and number of parameters of the respective decoder. PVTV2-b2 encoder has 3.91G FLOPS and  24.86M parameters. Small MERIT encoder has 24.62G FLOPs and 129.38M parameters. All the results are averaged over five runs.}
\label{tab:gcascade_aggregation_results}
\end{table}

\subsection{Comparison with the baseline decoder}

Table \ref{tab:compare_baseline_decoder} reports the experimental results with the computational complexity of our baseline CASCADE decoder and our proposed G-CASCADE decoder. We also report the results of original UpConv used in the CASCADE decoder and our modified efficient UCB. From Table \ref{tab:compare_baseline_decoder}, we can see that our modified UCB performs equal or better with significantly lower FLOPs and parameters. Our G-CASCADE decoder provides 0.5\% better DICE score than the CASCADE decoder with 80.8\% fewer parameters and 82.3\% fewer FLOPs.        

\subsection{Effect of different skip-aggregations in G-CASCADE decoder}
We conduct some experiments to see the effect of Additive or Concatenation in aggregating upsampled features with the skip-connections. Table \ref{tab:gcascade_aggregation_results} presents the results of PVT-GCASCADE and MERIT-GCASCADE with Additive and concatenation aggregations. We can see from Table \ref{tab:gcascade_aggregation_results} that Concatenation-based aggregation achieves marginally better DICE scores than Additive aggregation, while having significantly higher FLOPs and parameters. The reason behind this increase in computational complexity is the use of GCAM with the concatenated channels (i.e., 2$\times$ of original channels). Considering the lower computational complexity of Additive aggregation, we have used Additive aggregation in all of our experiments. 

\section{Conclusion}
\label{sec:conclusion}
In this paper, we have introduced a new graph-based cascaded convolutional attention decoder namely G-CASCADE for multi-stage feature aggregation. G-CASCADE enhances feature maps while preserving long-range information captured by transformers which is crucial for accurate medical image segmentation. Due to using graph convolution block instead of $3\times3$ convolution block, G-CASCADE is computationally efficient. Our experimental results show that G-CASCADE outperforms a recent decoder, CASCADE, in DICE score \color{black}with 80.8\% fewer parameters and 82.3\% fewer FLOPs\color{black}. Our experimental results also demonstrate the superiority of our G-CASCADE decoder over SOTA methods on \color{black}five \color{black}public medical image segmentation benchmarks. Finally, we believe that our proposed decoder will improve other downstream medical image segmentation and semantic segmentation tasks.





{\small
\bibliographystyle{ieee_fullname}
\bibliography{egpaper}
}

\appendix

\begin{table}[]
\begin{center}
    {\small{
\begin{tabular}{lrrrrrrrrrr}
\toprule
\multirow{2}{*}{Methods}        & \multicolumn{2}{c}{Avg} \\
                               & DICE         & mIoU           \\
\midrule
UNet \cite{ronneberger2015u} & 85.5       & 78.5          \\
UNet++ \cite{zhou2018unet++} & 80.9      & 72.9          \\
PraNet \cite{fan2020pranet} & 87.5       & 78.7       \\
CaraNet \cite{lou2022caranet} & 87.0      & 78.2      \\
TransUNet \cite{chen2021transunet}                   & 88.0       & 80.9       \\
TransFuse \cite{zhang2021transfuse}                   & 90.1       & 84.0       \\
UCTransNet \cite{wang2022uctransnet}                   & 90.5       & 83.0       \\
PolypPVT \cite{dong2021polyp}                      & 91.3       & 85.2       \\
PVT-CASCADE  \cite{Rahman_2023_WACV}                 &  91.1       & 84.9     \\
\midrule
PVT-GCASCADE (\textbf{Ours})                   &  \textbf{91.51$\pm$0.61}       & \textbf{86.53$\pm$0.54}     \\

\bottomrule 
\end{tabular}
}}
\end{center}
\caption{Results on ISIC2018 dataset. The results of UNet, UNet++, PraNet, CaraNet, TransUNet, TransFuse, UCTransNet, and PolypPVT are taken from \cite{tang2022duat}. We produce the results of PVT-CASCADE using our experimental settings for this dataset. All PVT-GCASCADE results are averaged over five runs. The best results are in bold.} 
\label{tab:isic2018_results}
\end{table}

\begin{table*}[]
\begin{center}
    {\small{
\begin{tabular}{lrrrrrrrrr}
\toprule
\multirow{2}{*}{Methods}        & \multicolumn{2}{c}{CVC-ClinicDB} & \multicolumn{2}{c}{Kvasir} & \multicolumn{2}{c}{ColonDB} & \multicolumn{2}{c}{EndoScene} \\
                               & DICE         & mIoU         & DICE            & mIoU           & DICE         & mIoU        & DICE         & mIoU             \\
\midrule
UNet \cite{ronneberger2015u}     & 82.3          & 75.5         & 81.8       & 74.6      & 51.2       & 44.4 & 71.0       & 62.7            \\
UNet++ \cite{zhou2018unet++}    & 79.4         & 72.9          & 82.1      & 74.3      & 48.3       & 41.0  & 70.7      & 62.4         \\
PraNet \cite{fan2020pranet}    & 89.9         & 84.9         & 89.8       & 84.0      & 71.2      & 64.0    & 87.1       & 79.7       \\
CaraNet \cite{lou2022caranet}    & 93.6         & 88.7         & 91.8       & 86.5      & 77.3      & 68.9  & 90.3       & 83.8      \\
UACANet-L \cite{kim2021uacanet} & 91.07      & 86.7         & 90.83       & 85.95      & 72.57      & 65.41   & 88.21      & 80.84         \\
SSFormerPVT \cite{wang2022stepwise}                   & 92.88          & 88.27         & 91.11       & 86.01      & 79.34       & 70.63     & 89.46       & 82.68          \\
PolypPVT \cite{dong2021polyp}                       & 93.08          & 88.28         & 91.23       & 86.3       & 80.75       & 71.85     & 88.71       & 81.89          \\
PVT-CASCADE \cite{Rahman_2023_WACV}                   & 94.34          & 89.98         & 92.58       & 87.76      & 82.54       & 74.53     &  90.47       & 83.79        \\
\midrule
PVT-GCASCADE (\textbf{Ours})                   &  \textbf{94.68}          & \textbf{90.18}         & \textbf{92.74}       & \textbf{87.90}      & \textbf{82.61}       & \textbf{74.60}     & \textbf{90.56}       & \textbf{83.87}      \\
\bottomrule 
\end{tabular}
}}
\end{center}
\caption{Results on polyp segmentation datasets. Training on combined Kvasir \cite{jha2020kvasir} and CVC-ClinicDB \cite{bernal2015wm} trainset. The results of UNet, UNet++ and PraNet are taken from \cite{fan2020pranet}. We get the results of PolypPVT, SSFormerPVT, and UACANet from \cite{Rahman_2023_WACV}. PVT-GCASCADE results are averaged over five runs. The best results are shown in bold.} 
\label{tab:polyp_results}
\end{table*}

\section{Datasets}

\textbf{ISIC2018 dataset.}
ISIC2018 dataset is a skin lesion segmentation dataset \cite{codella2019skin}. It consists of 2596 images with corresponding annotations. In our experiments, we resize the images to 384 $\times$ 384 resolution unless otherwise mentioned. We randomly split the images into 80\% for training, 10\% for validation,
and 10\% for testing.

\textbf{Polyp datasets.} Kvasir contains 1,000 polyp images collected from the polyp class in the Kvasir-SEG dataset \cite{jha2020kvasir}. CVC-ClinicDB \cite{bernal2015wm} consists of 612 images extracted from 31 colonoscopy videos.  Following CASCADE \cite{Rahman_2023_WACV}, we adopt the same 900 and 550 images from Kvasir and CVC-ClinicDB, respectively as the training set. We use the remaining 100 and 62 images as the respective testsets. To assess the generalizability of our proposed decoder, we use two unseen test datasets, namely EndoScene \cite{vazquez2017benchmark}, and ColonDB \cite{tajbakhsh2015automated}. EndoScene and ColonDB consists of 60 and 380 images, respectively.

\color{black}

\textbf{Retinal vessels segmentation datasets.} The DRIVE \cite{staal2004ridge} dataset has 40 retinal images with segmentation annotations. All the retinal images in this dataset are 8-bit color images of resolution $565 \times 584$ pixels. The official splits contain a training set of 20 images and a test set of 20 images. The CHASE\_DB1 \cite{carballal2018automatic} dataset contains 28 color retina images of $999 \times 960$ pixels resolution. There are two manual annotations of each image for segmentation. We use the first annotation as the ground truth. Following \cite{liu2022full}, we use the first 20 images for training, and the remaining 8 images for testing.

\color{black}

\section{Experiments}
\label{asec:experiments}

\color{black}
\subsection{Implementation details and evaluation metrics}
In this subsection, we discuss the implementation details of our proposed decoder for Retinal vessel segmentation.
We have conducted experiments on two retinal datasets such as DRIVE \cite{staal2004ridge} and CHASE\_DB1 \cite{carballal2018automatic}. In both cases, we first extend the training set using horizontal flips, vertical flips, horizontal-vertical flips, random rotations, random colors, and random Gaussian blurs. Through this process, we get 260 images including our 20 original training images. We use 26 of these images for validation that belong to 4 randomly selected original images. In the case of the DRIVE dataset, we resize the images into $768\times768$ resolution for PVT and ($768\times768$, $672\times672$) resolutions for MERIT. In the case of CHASE\_DB1, we use $960\times960$ resolution inputs for PVT and ($768\times768$, $672\times672$) resolution inputs for MERIT. However, we resize the output segmentation maps to the original resolution to get evaluation metrics during inference. We use random flips and rotations with a probability of 0.5 as augmentation methods during training. To train our models, we use the AdamW optimizer with both learning rate and weight decay of 1e-4. We optimize the combined weighted BCE and weighted mIoU loss function. The MUTATION is used to aggregate multi-stage loss. We train our networks for 200 epochs with a batch size of 4 and 2 for DRIVE and CHASE\_DB, respectively.  

We use accuracy (Acc), sensitivity (Sen), specificity (Sp), DICE, and IoU scores as evaluation metrics. We report the percentage (\%) score averaging over five runs for both datasets. 
\color{black}

\subsection{Experimental results on ISIC2018 dataset}

Table \ref{tab:isic2018_results}
presents the average DICE scores of our PVT-GCASCADE and MERIT-GCASCADE along with other SOTA methods on the ISIC2018 dataset. This dataset is different than the CT and MRI images used in the above experiments. In this case also, it is evident from the table that our PVT-GCASCADE achieves the best average DICE (91.51\%) and mIoU (86.53\%) scores. PVT-GCASCADE outperforms its counterpart PVT-CASCADE by 0.4\% DICE and 0.6\% mIoU scores.   


\begin{table}[t]
\color{black}
\centering 
    {
\begin{adjustbox}{width=0.48\textwidth}
{\begin{tabular}{lrrrrrr}
\toprule
Methods      & Acc & Sen & Sp & DICE & IoU \\
\midrule
UNet \cite{ronneberger2015u}                    &  96.78 &  80.57 & 98.33 & 81.41 &   68.64 \\
UNet++ \cite{zhou2018unet++}                    &  96.79 &  78.91 & \textbf{98.50} & 81.14 &   68.27 \\
Attention UNet \cite{oktay2018attention}                    &  96.62 &  79.06 & 98.31 & 80.39 &   67.21 \\
FR-UNet \cite{liu2022full}                    &  97.05 &  \textbf{83.56} & 98.37 & \textbf{83.16} &   \textbf{71.20} \\
PVTV2-b2 (only) \cite{wang2022pvt}                    & 96.24 & 82.02 & 97.61 & 79.14 & 65.48  \\
PVT-CASCADE \cite{Rahman_2023_WACV}                   & 96.79 & 83.07 & 98.10 & 81.73 &  69.10 \\

MERIT-CASCADE \cite{rahman2023multi}                              & 96.89 &  82.94 & 98.22 & 82.21 & 69.08   \\
\midrule 
PVT-GCASCADE (\textbf{Ours})                     & 96.89 & 83.00 & 98.22 & 82.10 &   69.70    \\
MERIT-GCASCADE (\textbf{Ours})                & \textbf{97.07} & 82.81     & 98.44 & 82.90 &   70.81 \\
\bottomrule \\
\end{tabular}}
\end{adjustbox}

}
\caption{Results (\%) of Retinal Vessel Segmentation on DRIVE dataset. The results of UNet, UNet++, Attention UNet, and FR-UNet are taken from \cite{liu2022full}. All other results are averaged over five runs in our experimental setups. The best results are in bold.}
\label{tab:results_drive}
\color{black}
\end{table}

\begin{table}[t]
\color{black}
\centering 
    {
\begin{adjustbox}{width=0.48\textwidth}
{\begin{tabular}{lrrrrrr}
\toprule
Methods      &  Acc & Sen & Sp & DICE & IoU \\
\midrule
UNet \cite{ronneberger2015u}                    &  97.43 &  76.50 & 98.84 & 78.98 &   65.26 \\
UNet++ \cite{zhou2018unet++}                    &  97.39 &  83.57 & 98.32 & 80.15 &   66.88 \\
Attention UNet \cite{oktay2018attention}                    &  97.30 &  83.84 & 98.20 & 79.64 &   66.17 \\
FR-UNet \cite{liu2022full}                   &  97.48 &  \textbf{87.98} & 98.14 & 81.51 &   68.82 \\
PVTV2-b2 (only) \cite{wang2022pvt}                   & 97.25 & 85.07 & 98.07 & 79.58 &  66.12 \\
PVT-CASCADE \cite{Rahman_2023_WACV}                   & 97.55 & 85.83 & 98.34 & 81.50 &  68.80 \\
MERIT-CASCADE \cite{rahman2023multi}                             &  97.60 & 84.97 & 98.45 & 81.68 &   69.06 \\
\midrule 
PVT-GCASCADE (\textbf{Ours})                       & 97.71 & 85.84 & 98.51 & 82.51 & 70.24      \\
MERIT-GCASCADE (\textbf{Ours})             &  \textbf{97.76} & 84.93     & \textbf{98.62} & \textbf{82.67} &    \textbf{70.50} \\
\bottomrule \\
\end{tabular}}
\end{adjustbox}

}
\caption{Results (\%) of Retinal Vessel Segmentation on CHASE\_DB1 dataset. The results of UNet, UNet++, Attention UNet, and FR-UNet are taken from \cite{liu2022full}. All other results are averaged over five runs in our experimental setups. The best results are in bold.}
\label{tab:results_chase}
\color{black}
\end{table}

\subsection{Experimental results on Polyp datasets}
We evaluate the performance and generalizability of our G-CASCADE decoder on four different polyp segmentation testsets among which two are completely unseen datasets collected from different labs. Table \ref{tab:polyp_results} displays the DICE and mIoU scores of SOTA methods along with our G-CASCADE decoder. From Table \ref{tab:polyp_results}, we can see that G-CASCADE significantly outperforms all other methods on both DICE and mIoU scores. 
It is noteworthy that G-CASCADE outperforms the best CNN-based model UACANet by a large margin on unseen datasets (i.e., 9.8\% DICE score improvement in ColonDB). Therefore, we can conclude that due to using transformers as a backbone network and our graph-based convolutional attention decoder, PVT-GCASCADE inherits the merits of transformers, GCNs, CNNs, and local attention which makes them highly generalizable for unseen datasets.

\color{black}
\subsection{Experimental results on Retinal vessels segmentation datasets}

We have conducted experiments on two retinal vessel segmentation datasets such as DRIVE and CHASE\_DB1. The experimental results are reported in Tables \ref{tab:results_drive} and \ref{tab:results_chase}. Our G-CASCADE decoder outperforms the baseline CASCADE decoder with significantly lower computational costs. Specifically, our PVT-GCASCADE shows 0.37\% and 1.01\% improvements in DICE score over PVT-CASCADE in DRIVE and CHASE\_DB1 datasets, respectively. Similarly, our MERIT-GCASCADE exhibits 0.69\% and 0.99\% improvements in DICE score in DRIVE and CHASE\_DB1 datasets, respectively. From Tables \ref{tab:results_drive} and \ref{tab:results_chase}, we can conclude that our methods show competitive performance compared to the SOTA approaches. Although FR-UNet achieves a 0.26\% better DICE score in the DRIVE dataset, it has a 1.16\% lower DICE score in CHASE\_DB1 than our MERIT-GCASCADE. Besides, FR-UNet splits the retinal images into $48\times48$ pixels patches in a stride of 6 pixels during training but we use the whole retinal images during both training and inference. Consequently, we have a significantly lower number of samples for training compared to FR-UNet. We can conclude from the results that our G-CASCADE decoder equally performs well in retinal vessel segmentation.

\color{black}

\begin{table}[]
\centering 
    {
\begin{adjustbox}{width=0.48\textwidth}
{\begin{tabular}{lrrrr}
\toprule
Graph Convolutions   &  \multicolumn{1}{c}{FLOPs(G)} & \multicolumn{1}{c}{\#Params(M)} & \multicolumn{1}{l}{DICE (\%)} \\
\midrule
GIN \cite{xu2018powerful} & 0.313 & 1.59 & 82.22  \\
EdgeConv \cite{wang2019dynamic} & 0.957 & 1.78 & 82.81   \\
GraphSAGE \cite{hamilton2017inductive} & 0.520 & 1.88 & 83.10      \\
Max-Relative \cite{li2019deepgcns} (\textbf{Ours}) & 0.342 & 1.78 & \textbf{83.28}   \\
\bottomrule \\
\end{tabular}}
\end{adjustbox}

}
\caption{Experimental results of different graph convolutions in GCAM block on Synapse Multi-organ dataset. We use the PVTV2-b2 encoder and only report the FLOPs and number of parameters of the decoder. All the results are averaged over five runs. The best results are shown in bold.}
\label{tab:different_graph_convs}
\end{table}

\begin{table}[t]
\color{black}
\centering 
    {
\begin{adjustbox}{width=0.48\textwidth}
{\begin{tabular}{lrrrr}
\toprule
Architectures      &  \multicolumn{1}{c}{FLOPs(G)} & \multicolumn{1}{c}{\#Params(M)} & \multicolumn{1}{l}{DICE (\%)} \\
\midrule
PVT-CASCADE                       & 5.84 & 34.13 & 83.28  \\
PVT-GCASCADE                        & \textbf{4.252} & \textbf{26.64} & \textbf{83.40}      \\
\midrule 
MERIT-CASCADE                                 & 33.31 & 147.86 & 84.54   \\
MERIT-GCASCADE                   & \textbf{26.143} & \textbf{132.93}     & \textbf{84.63}   \\
\bottomrule \\
\end{tabular}}
\end{adjustbox}

}
\caption{Comparison of overall computational complexity. We use the PVTV2-b2 backbone with an input resolution of $224\times224$ in both PVT-CASCADE and PVT-GCASCADE. We use two Small MaxViT backbones with input resolutions of $256\times256$ and $224\times224$ in MERIT architectures.}
\label{tab:overall_computational_complexity_comparision}
\color{black}
\end{table}

\begin{table}[]
\centering 
    {
{\begin{tabular}{lrrrr}
\toprule
Input resolutions   & \multicolumn{1}{l}{DICE (\%)} & \multicolumn{1}{l}{mIoU (\%)} & \multicolumn{1}{l}{HD95 (\%)} \\
\midrule
224$\times$224 & 83.28  &  73.91 & 15.83\\
256$\times$256 & 84.21  &  75.32 & 14.58 \\
384$\times$384 & 86.01   & 78.10  &  13.67 \\
\bottomrule \\
\end{tabular}}

}
\caption{Experimental results of PVT-GCASCADE with different input resolutions on Synapse Multi-organ dataset. All the results are averaged over five runs.}
\label{tab:different_input_resolutions}
\end{table}

\section{Ablation Study}
\subsection{Comparison among different graph convolutions in GCAM}
We report the experimental results of our decoder with different graph convolutions in Table \ref{tab:different_graph_convs}. As shown in Table \ref{tab:different_graph_convs}, Max-Relative (MR) \cite{li2019deepgcns} graph convolution provides the best DICE score (83.28\%) with only 0.342G FLOPs and 1.78M parameters. Although GIN \cite{xu2018powerful} has slightly lower FLOPs and parameters, it provides the lowest DICE score (82.22\%).
EdgeConv \cite{wang2019dynamic} and GraphSAGE \cite{hamilton2017inductive} graph convolutions have lower DICE scores than the MR graph convolution with higher computational costs.

\color{black}
\subsection{Overall computational complexity}
We report the total parameters and FLOPs of encoder backbones and our decoder in Table \ref{tab:overall_computational_complexity_comparision}. We can see from Table \ref{tab:overall_computational_complexity_comparision} that overall computational complexity depends on the number of parameters and FLOPs of the encoder backbones. We implement our decoder on top of PVTV2-b2 \cite{wang2022pvt} and Small MaxViT \cite{tu2022maxvit} backbones. Our PVT-GCASCADE has 4.252G FLOPs and 26.64M parameters, which is 1.588G and 7.49M lower than the corresponding PVT-CASCADE architecture. Due to the larger size of two Small MaxViT backbones in MERIT-CASCADE architecture (i.e., 33.31G FLOPs and 147.86M parameters), our MERIT-GCASCADE (i.e., 26.143G FLOPs and 132.93M parameters) is also larger in size. In both cases, the savings in FLOPs and parameters come only from our decoder. Our proposed decoder can easily be plugged into other hierarchical encoders; if a lightweight encoder is used, the total computational cost will be reduced.
\color{black}

\subsection{Influence of input resolution}

Table \ref{tab:different_input_resolutions} presents the quantitative segmentation performance of PVT-GCASCADE network with different input resolutions. We conduct experiments with three input resolutions such as 224$\times$224, 256$\times$256, and 384$\times$384. It is evident from the table that performance improved in all three evaluation metrics for higher input resolutions. We get the best DICE and mIoU  86.01\% and 78.10\%, respectively with the input resolution of 384$\times$384.  

\end{document}